\documentclass{mnras}
\usepackage{graphicx}

\def\pg{{PG1211+143}}
\def\mkn{{Mkn509}}

\def\mkn{{Mkn509}}

\def\xmm{{\it XMM-Newton}}

\def\suzaku{{\it Suzaku}}

\def\et{{et al.\ }}

\newcommand{\ls}{\mathrel{\hbox{\rlap{\hbox{\lower4pt\hbox{$\sim$}}}\hbox{$<$}}}}
\newcommand{\gs}{\mathrel{\hbox{\rlap{\hbox{\lower4pt\hbox{$\sim$}}}\hbox{$>$}}}}

\def\Msun{\hbox{$\rm ~M_{\odot}$}}

\def\H0{{\rm ~km~s^{-1}~Mpc^{-1}}}

\def\et{{et al.}}

\title [Slow inflow]
       {Low-redshift absorption in the Seyfert galaxy \pg\ - a distant inflow maintaining off-plane accretion or the gravitational redshift of matter orbiting the SMBH?}
 \author[Ken Pounds and Kim Page]
       {Ken Pounds and Kim Page\\
School of Physics and Astronomy, University of Leicester, Leicester, LE1 7RH, UK\\}

\date{Accepted ; Submitted }
\pagerange{\pageref{firstpage}--\pageref{lastpage}}
\pubyear{2023}
\begin{document}

\label{firstpage}
\pagerange{\pageref{firstpage}--\pageref{lastpage}}

\maketitle

\begin{abstract}
The detection of a high velocity ($\sim$ 0.3c) {\it inflow} of highly ionized matter during an extended \xmm\ observation of the luminous Seyfert
galaxy \pg\ in 2014 provided the first direct evidence of a short-lived accretion event, and an explanation for the powerful winds (UFOs) now recognised
as a common property of many luminous Seyfert galaxies. Although the ultra-fast inflow - observed at a redshift of 0.483 - was detected in only one of seven spacecraft orbits,  
weaker (lower column) but more persistent absorption is seen - at a redshift of 0.123 - in the high exposure, soft X-ray (RGS) spectra summed over all 7 individual spacecraft orbits.
Similar stacking of the higher energy (pn camera) spectra reveals underlying absorption at a redshift of 0.148.
Interpreted, conventionally, as a Doppler redshift, the RGS observation indicates a line-of-sight inflow velocity v$\sim$ 0.038c and (free-fall) radial location at 1400 R$_{g}$, with the higher redshift and
ionization in pn camera spectra perhaps detecting that inflow closer to the black hole.
A very different explanation would be absorption in matter subject to the strong gravity close to the SMBH, an interpretation supported by the launch of a new UFO in the final spacecraft
orbit.
\end{abstract}


\section{Introduction}
\xmm\ observations of the luminous Seyfert galaxy \pg\ have been remarkably productive, since chosen as a target in the `guaranteed observing
time' awarded to the late Martin Turner,
a PI of the EPIC focal plane camera (Turner \et\ 2001). An early observation showed a strong, blue-shifted
absorption line of highly ionized Fe, interpreted as an outflow or wind, with a line-of-sight velocity of 0.15$\pm$0.01$c$ (Pounds \et\ 2003; Pounds and
Page 2006). Archival data from \xmm\
and the Japanese X-ray observatory \suzaku\ subsequently showed that such ultra-fast,
highly-ionized outflows (UFOs) are relatively common in nearby, luminous AGN (Tombesi \et\ 2010, 2011; Gofford \et\ 2013).

While those archival
searches typically reported a single velocity, in the few cases where an AGN was observed repeatedly the wind velocity was often different, with
\mkn\ being the best example from the \xmm\ data archive, with wind velocities of $\sim$0.173c, $\sim$0.139c and $\sim$0.196c, separated by 5 years
and 6 months respectively (Cappi \et\ 2009). Looking back, it is interesting to note that while such winds were predicted to occur in luminous AGN when a
region of the accretion disc
became `super-Eddington' (Shakura and Sunyaev 1973; SS73), they were not part of the original scientific case for the X-ray Multi-Mirror
Observatory (Bleeker \et\ 1984).

An extended \xmm\ observation of \pg\ in 2014, covering seven consecutive spacecraft orbits over five weeks, showed a more complex wind structure, with
three primary (high column) outflow velocities of $v \sim 0.066c$, $v \sim 0.129c$ and $v \sim 0.187c$, observed in EPIC pn
(Pounds \et\ 2016a; P16a) and RGS spectra (Pounds \et\ 2016b; P16b), none being consistent with the velocity observed in 2001.  In a review of continuum-driven or
`Eddington' winds, King and Pounds (2015) furthermore showed that the observability of an individual, short-lived wind ejection may be of order months or
less, as an expanding shell presents a lower absorbing column to the nuclear X-ray source.

Considering the simultaneous observation of multiple expanding shells of absorbing gas in the context of super-Eddington accretion,  Pounds,
Lobban  and Nixon (2017) noted that short-lived winds might be a natural  consequence of the way in which matter accretes in an AGN between mergers,
typically falling from far outside the radius of gravitational influence of the SMBH and with essentially random orientation. The resulting 
accretion stream will in general orbit in a plane misaligned to the spin of the central black hole (King and Pringle 2006, 2007), with the inner
disc subject to Lense-Thirring precession around the spin vector, and orbits at smaller radii precessing sufficiently fast to cause
the tearing-away of independent rings of gas.

Computer simulations by Nixon \et\ (2012) show that - as each torn-off ring precesses on its own timescale - two neighbouring rings will eventually
collide, with the shocked material losing rotational support and falling inwards to a new radius defined by its residual angular momentum. A substantial
increase in the local accretion rate might then become briefly super-Eddington, with excess matter being ejected as a wind with
velocity at or above the local escape velocity (as predicted in SS73).
For typical AGN disc parameters, Nixon \et\ (2012) find the `tearing radius' to be of
order a few hundred gravitational radii from the black hole - with predicted (escape) wind speeds in the range observed (King and Pounds 2015;
fig 4a).

An important bonus from the 2014 study of \pg\ came with the detection of red-shifted X-ray absorption spectra during the second \xmm\ orbit (rev
2659), revealing a stream of matter approaching the black hole at velocities up to
$\sim$ 0.3c (Pounds \et\ 2018) and offering the first observational support for the random accretion scenario. Simultaneous hard and soft X-ray spectra
provided independent confirmation of the 0.48 redshift, with the soft X-ray absorption indicating
less ionized (higher density) matter embedded in a more highly ionized, more massive inflow. A stronger (higher column) soft X-ray absorption
component was also detected, at a redshift of 0.189, corresponding to a line-of-sight velocity of $\sim$0.1c, and `upstream' location at 200 R$_{g}$. We now report more persistent absorption, at
lower redshifts of 0.123 and 0.148, in the high exposure spectra obtained by stacking data from the RGS (den Herder \et\ 2001) and pn detector (Strueder \et\ 2001)
over the full 7-orbit observation. Data stacking was carried out with the \xmm\ analysis tools RGSCOMBINE and MATHPHA (https:xmm-tools.cosmos.esa.int).  

We assume a cosmological redshift for \pg\ of $z=0.0809$ (Marziani \et\ 1996), with a black hole mass of $4\times 10^{7}$\Msun\ (Kaspi \et\
2000) confirming an historical bolometric luminosity close to Eddington. Spectral modelling is based on the {\tt XSTAR} software (Arnaud 1996) and includes absorption due to the line-of sight
Galactic column  N$_{H}$ $\sim$ 3$\times$ 10$^{20}$ cm$^{-2}$ (Murphy \et\ 1996).

For the pn data we follow the procedure described in P16a, where discrete spectral features are compared with publicly available grids of pre-computed photoionized absorption
and emission spectra based on the {\tt XSTAR} v2.2 code (Kallman \et\ 1996) and the ionizing continuum is modelled by a double power law.
As noted in P16b, those publicly available grids failed to fit a strong Fe UTA (unresolved transition array) at $\sim$16--17 \AA\ in RGS spectra without an unrealistically high Fe abundance.
For the RGS analysis
we therefore generated a self-consistent set of multiplicative absorption and additive emission grids, also used for the RGS analysis here, with the spectral energy distribution (SED) of \pg\ based  on
concurrent data from the Optical Monitor (Mason \et 2001) and EPIC-pn cameras, extrapolated from 1-1000 Rydberg. The SED is described in more detail in Lobban \et (2016).

Free parameters in modelling the absorption spectra are column density, ionization  parameter, and observed blue/red-shift. For the emission spectra the column density is
fixed at a suitably low value ($N_{\rm H}$ = 10$^{20}$ cm$^{-2}$) to avoid significant opacity effects, with the ionization, flux level and line-of-sight velocity again as free parameters.

\section{Persistent low-redshift absorption in the stacked RGS spectrum}
In the initial spectral analysis of the 2014 RGS data, Pounds \et (P16b) report a mean soft X-ray spectrum, with an array
of photoionized emission and absorption features superposed on the X-ray continuum modelled by hard and soft power laws. In modelling the transient absorption
event identified with an ultrafast inflow in the second \xmm\ orbit, the continuum components
and photoionised emission spectra were retained, with only normalisations left free (Pounds \et\ 2018). Parameters of the blue-shifted absorption attributed to
outflowing matter were also retained, the primary aim being to model and quantify the transient ultrafast inflow. We follow a similar procedure here to search for weaker
but more persistent absorption features in the stacked RGS1 and RGS2 data, with a high combined on-source exposure of 1.24 Ms. 

Two additional low-redshift absorption components are found:\\
(1) at a redshift of 0.089$\pm$ 0.001, with ionization parameter log $\xi$ $\sim$ 0.83 and column density N$_{H}$ $\sim$ 3$\times$ 10$^{19}$ cm$^{-2}$; and \\
(2) at a redshift of 0.123$\pm$ 0.001, with ionization parameter log $\xi$ $\sim$ 2.1 and column density N$_{H}$ $\sim$ 2.5$\times$ 10$^{20}$ cm$^{-2}$

\begin{figure}                                 
\centering                                                              
\includegraphics[width=10cm]{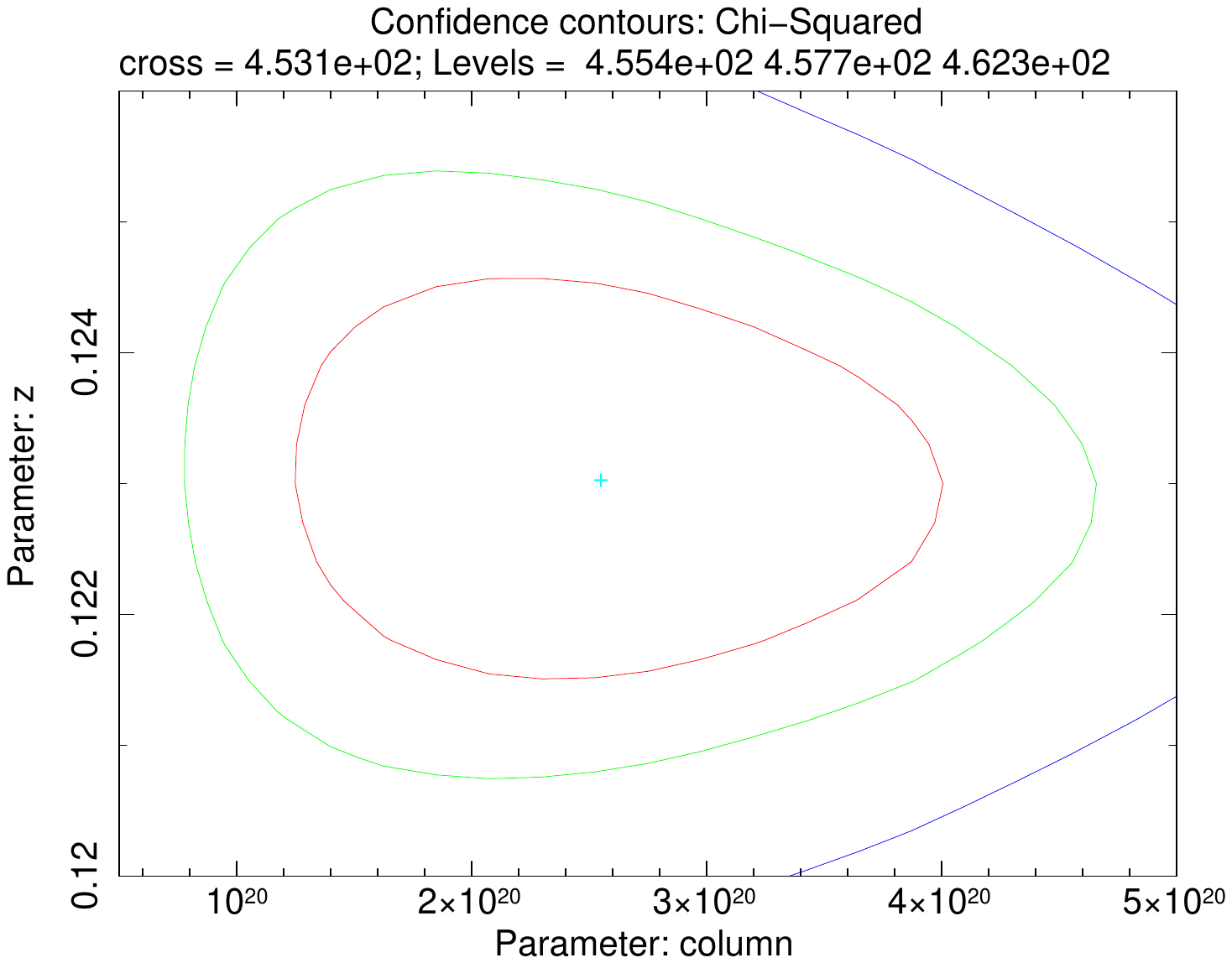}
\includegraphics[width=10cm]{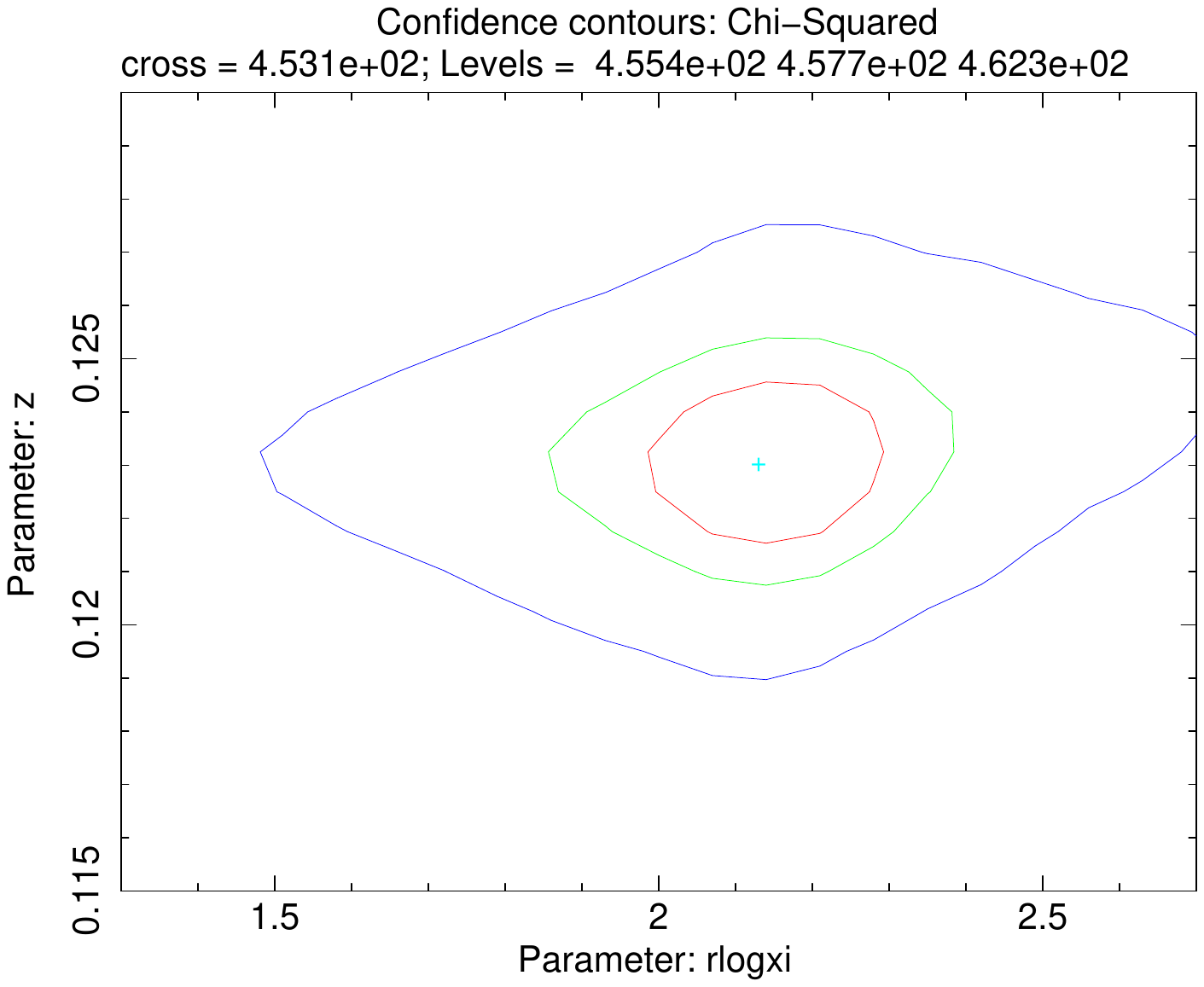}
\caption{Confidence contour maps of the low redshift absorber detected in the stacked RGS data from the 2014 \xmm\ observation of \pg.
Contours represent 1,2 and 3 sigma limits for the redshift as functions of column density (top) and ionization parameter (lower)} 
\end{figure}

The first component has a redshift close to the cosmological value for \pg\ and low ionization parameter suggesting a small change to the adopted
values for the line-of-sight ISM in our spectral model.

More interesting is the component with observed redshift of $\sim$0.123. Allowing for the cosmological redshift and a relativistic Doppler correction the corresponding (AGN rest-frame) inflow velocity is
0.038$\pm$0.001c. Repeating the spectral fit without this component
showed a null probability of 0.03.

A further test showed weak evidence for soft X-ray absorption at a higher redshift of 0.147$\pm$0.002, though with a null probability by the standard f-test of 0.28. 

Figure 1 shows probability contour plots for the RGS redshift as a function of column
density and of ionization parameter.

\section{Evidence of similar low redshift absorption in the stacked pn spectrum}
The relatively high ionization parameter of the low redshift absorption in the RGS data suggested similar persistent absorption might also be seen in the stacked pn data, which extends to higher
photon energies. 

\begin{figure}                                 
\centering                                                              
\includegraphics[width=10cm]{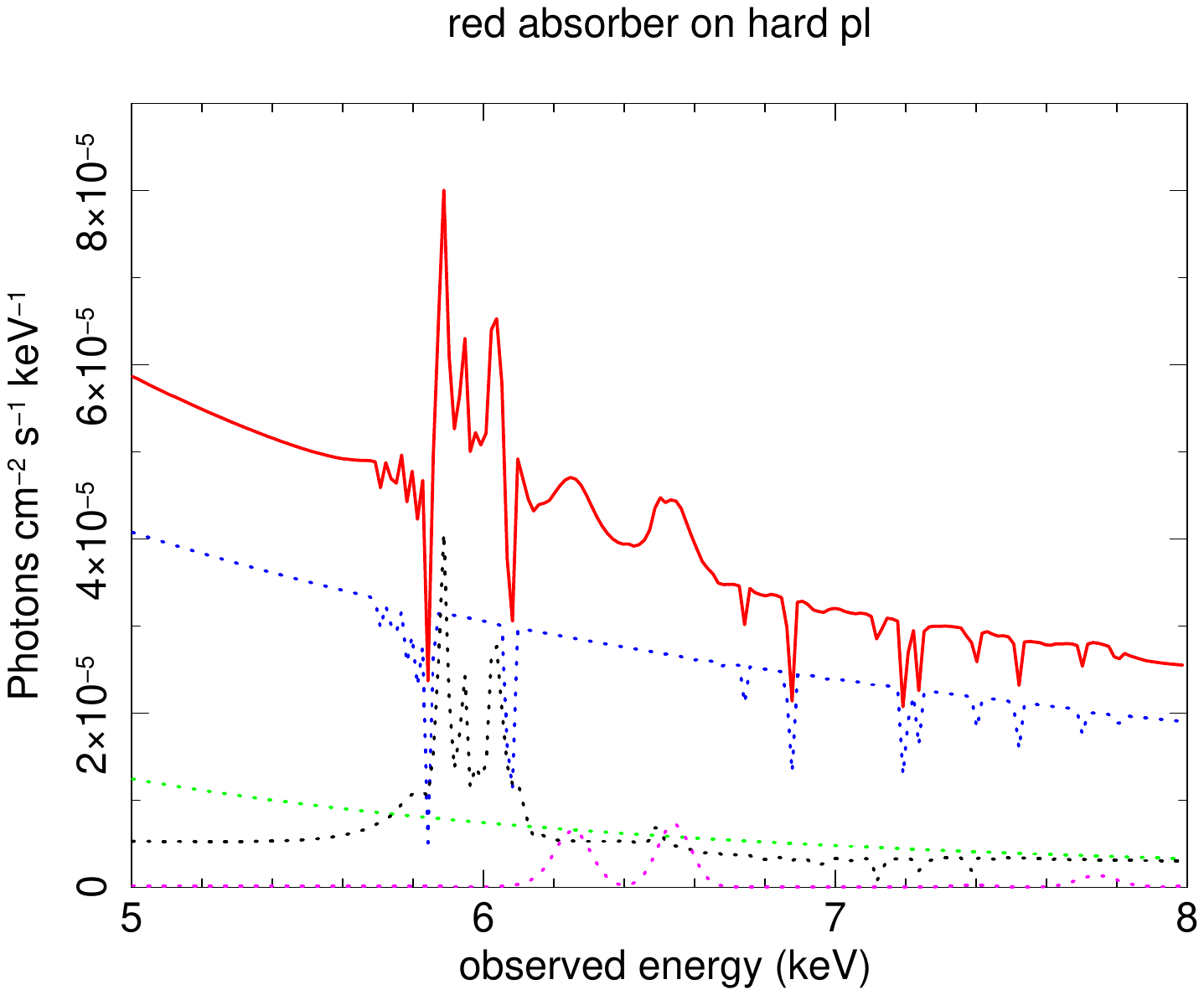}
\includegraphics[width=10cm]{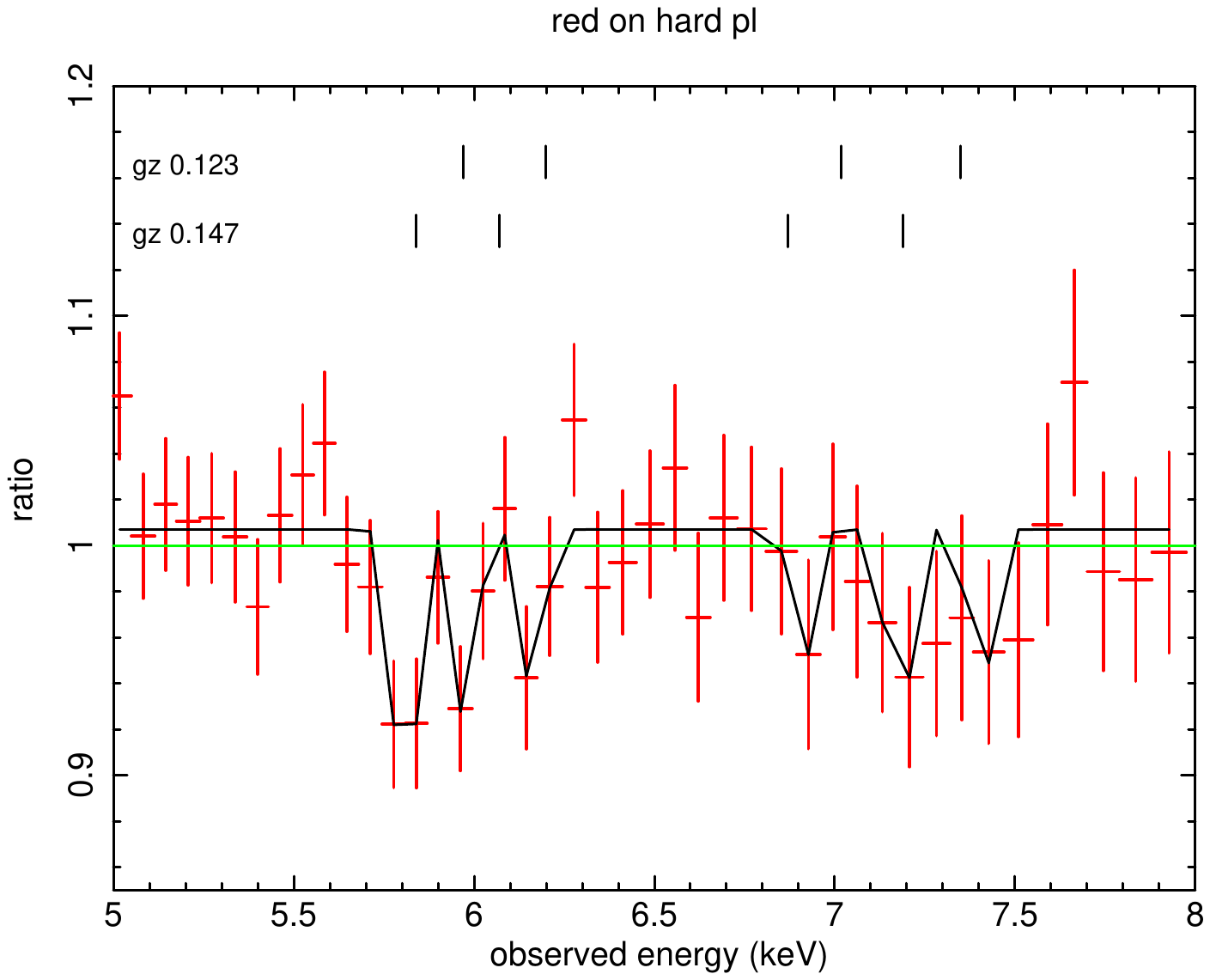}
\caption{(upper panel) Components of the photoionization model described in the text, with absorption at redshift 0.147 only on the hard power law continuum (shown in blue). Other spectral
components are a soft power law (green), reflection from the accretion disc (black) and photoionized emission from the outflow (magenta).
(lower panel) Ratio of data-to-model spectrum when the 0.147 redshift absorber is removed, revealing residual features identified with resonance 1s-2p and 1s-3p transitions in helium-like
and H-like Fe at a common redshift of 0.147. Not included in the photoionization model - but also seen in the ratio plot -  are absorption components consistent with the lower redshift of 0.123. as
discissed in the text.} 
\end{figure}

The spectral modelling of stacked pn camera data described in P16a was then repeated with the addition of a low redshift absorption component.

Assuming, initially, the additional absorption relates to a distant inflow, it was applied to the complete (continuum plus photo-ionized emission and reflection) spectrum, finding weak evidence for absorption 
at a redshift of 0.149$\pm$ 0.004, with ionization parameter log $\xi$=3.4$\pm$0.1 and line-of-sight column density N$_{H}$ = 3.2$\pm$1.7$\times 10^{22}$ cm$^{-2}$. 
Removal of the additional red-shifted absorber, and re-fitting with the hard power law index and all continuum emission normalisations free, increased the fit statistic by $\Delta \chi^2$
of 10/3 d.o.f  (for a null probability of 0.02).

To quantify the alternative, strong-gravity origin, the pn spectral fit was repeated, with the low redshift absorber now covering only the hard power law (or disc corona) emission and disc reflection.
Absorption was again detected, at a similar redshift of 0.147$\pm$ 0.003, with ionization parameter log $\xi$=3.3$\pm$0.1, and a higher column density N$_{H}$ = 8$\pm$3$\times 10^{22}$ cm$^{-2}$. 
Removal of the red-shifted absorber, and refitting with only the hard power law and reflection components free to change, was more highly significant with the fit statistic increasing by $\Delta \chi^2$
of 18/3 d.o.f, with a null probability of 3$\times$ 10$^{-4}$.

Figure 2 (upper panel) illustrates that second model, with the strong absorption line of He-like Fe (rest energy 6.70 keV) located at 5.84 keV. Satellite lines on the low energy wing -
seen in the model but not resolved by the pn solid state detector -
explains the small shift of the corresponding feature in the ratio plot.

Although not strongly required in the photoionization modelling, it is interesting to see separate components that correspond to a higher energy inflow at the same, lower, redshift dominating the
softer RGS spectra. With a re-examination of the stacked RGS data showing an additional, weaker, absorption component at redshift 0.146, we believe that the data show two separate inflows, a conclusion that
also explains why the identfied absorption features have width compatible with the pn spectral resolution, in turn providing an upper limit to the radial thickness of each ring, as noted below.

\begin{table*}
\centering
\caption{Parameters of the additional low-redshift absorber obtained from a 2--10 keV spectral fit to the stacked 2014 pn data, with the absorber applied (a) to the complete (continuum,
photoionized emission and reflection) spectrum; and (b) to the hard power law only. The absorber is defined by its ionization parameter $\xi$ (erg cm s$^{-1}$), column density $N_{\rm H}$ (cm$^{-2}$) and
redshift. Also shown for each case is the increase in $\chi^{2}$ when the redshifted absorber is removed and the spectrum re-fitted, together with the corresponding null probability.} 
\begin{tabular}{@{}lccccc@{}}
\hline
absorber geometry & log$\xi$ & $N_{\rm H}$($10^{23}$) & redshift & $\Delta \chi^{2}$ & random chance \\
\hline
(1)absorber on complete spectrum & 3.4$\pm$0.1 & 0.149$\pm$0.004  & 3.2$\pm$1.7$\times 10^{22}$ & 10/3 & 0.02 \\
(2)absorber on hard power law only & 3.4$\pm$0.1 & 0.147$\pm$0.003  & 8$\pm$3$\times 10^{22}$  & 18/3 & 0.0003 \\
\hline
\end{tabular}
\end{table*}

\section{Discussion}

The detection of a fast highly ionized wind in early X-ray observations of the narrow line Seyfert galaxy \pg\ (Pounds \et\ 2003) and
the luminous QSO PDS 456 (Reeves \et\ 2003) opened a new field of study of AGN, well matched to the high throughput of X-ray
spectrometers on ESA's \xmm, launched in late 1999. King and Pounds (2003) noted that such winds are a natural result of a high accretion ratio,
with excess matter being driven off the disc by radiation pressure when the accretion rate exceeds the local Eddington limit, as predicted in the classic review
of AGN accretion disc physics by Shakura and Sunyaev (1973). While that picture provides a satisfactory explanation of most UFOs, where a single detection yielded a unique wind velocity,
the extended study of \pg\ in 2014 found a more complex outflow velocity profile, suggesting some intrinsic disc instability or a rapidly variable
accretion rate. The detection of multiple red-shifted X-ray absorption lines during part of the same 2014 \xmm\ observation of \pg, where a common redshift
($\sim$0.48) indicated absorption in a substantial {\it inflow} approaching the black hole at a velocity of $\sim$ 0.3c (Pounds \et\ 2018), offered the first observational
support for such transient accretion.    

In the present paper we report more persistent red-shifted absorption in \pg, at a lower redshift, detected in high exposure spectra obtained by stacking RGS and pn camera data over the whole
5-weeks \xmm\ observation.
Here, we briefly consider two explanations for this new spectral component.

After correcting for the cosmological redshift of \pg, an intrinsic redshift of 0.042 in the RGS data might relate to 
a slow infall of ionized matter, with line-of-sight velocity of 0.038$\pm$0.001c and (free-fall) location at 1400 R$_{g}$ - beyond the tearing region discussed by Nixon \et\
(2012), but within the sphere of influence of the SMBH in \pg, perhaps having an important role in off-plane accretion on a multi-decadel timescale.

A very different interpretation would place the absorbing matter much closer to the SMBH, where a
redshift of 0.042 would correspond to the effect of strong gravity on matter orbiting the SMBH at 27 R$_{g}$.
While Lense-Thirring precession would move such a ring of matter across the line of sight in of order 3.5/a yr, where 0$<$a$<$1 is the Kerr spin parameter, that timescale is
readily consistent with an observed redshift persisting over the 5-weeks observation in 2014 - and perhaps still evident today - unless being rapidly accreted. In the latter context, the launch of a
powerful new UFO in the final days of the 2014 campaign (Pounds and Nayakshin) is of particular interest. 

Can a strong choice now be made for the relativistic Doppler
or strong gravity interpretations of the persistent redshift detected in the high-exposure stacked data from 2014?

At first sight the narrow soft x-ray absorption line spectrum appears incompatible with a strong gravity origin. Fig.3 shows a section of the stacked RGS soft X-ray spectrum, with the low
redshift absorber removed, revealing 6 K-shell absorption lines of O, Ne, and Mg, together with 2 of Fe-L, each consistent with a common redshift of 0.123. The lines are
plotted with Gaussians of 1$\sigma$ width 30 m\AA, consistent with the pre- and early post-launch RGS resolution over the relevant waveband (den Herder \et\ 2001), and show no obvious intrinsic broadening. 

To quantify that constraint, the set of 9 absorption lines was re-modelled with the addition of a one-$\sigma$ 'physical' line width, and
the quality of the fit determined as that common component was increased, from 20 m\AA\ to 80 m\AA\ in steps of 5 m\AA. The result was a strong upper limit to any physical line broadening
of 20 m\AA. 

\begin{figure*}                                 
\centering                                                              
\includegraphics[width=16cm]{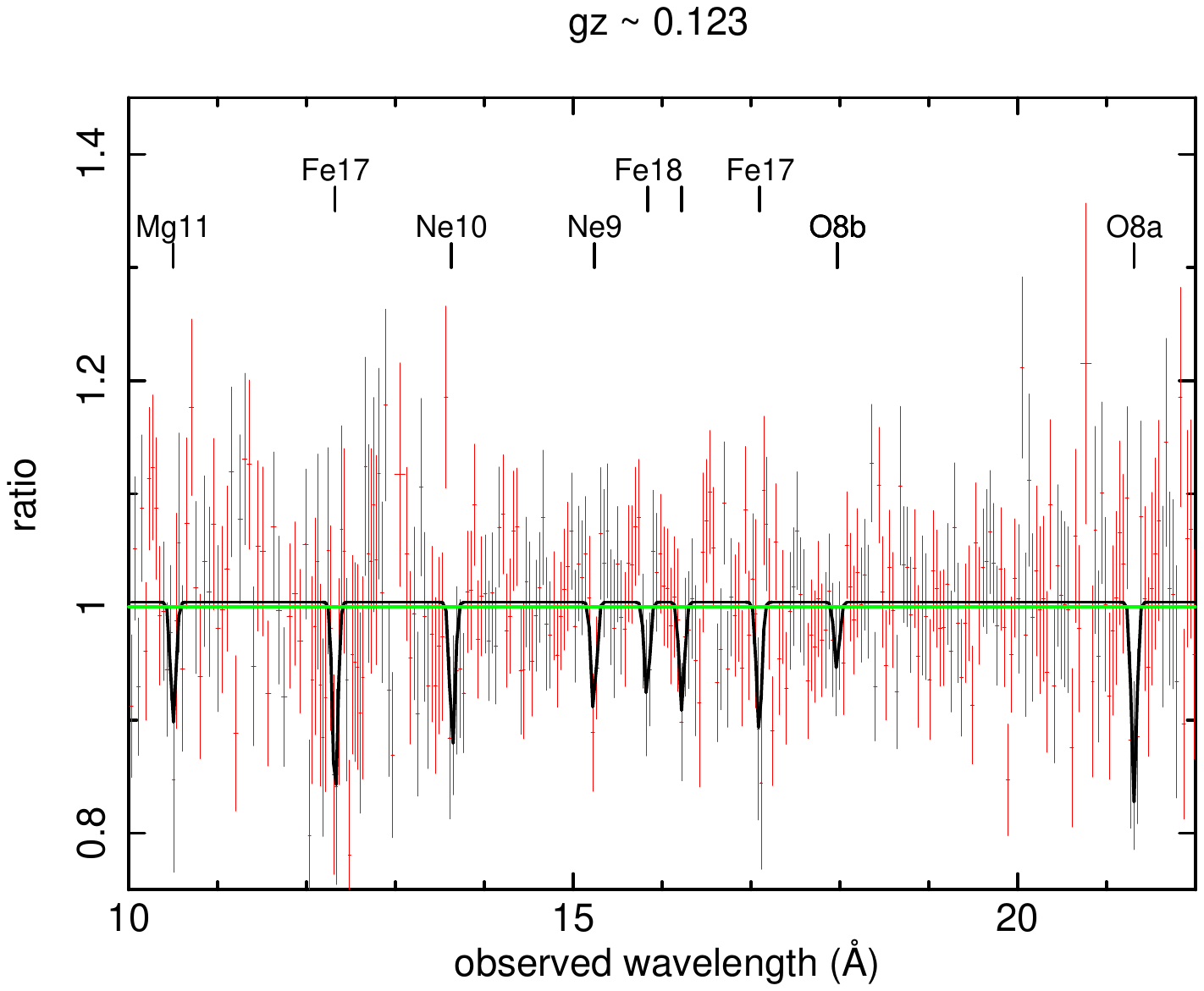}
\caption{RGS spectrum of the low redshift absorber detected in stacked data from the 2014 \xmm\ observation of \pg.
K=shell absorption lines of oxygen, neon, magnesium and iron are plotted with Gaussian profiles consistent with the pre-launch RGS resolution (den Herder \et\ 2001), with markers indicating the
expected wavelength of each identified line at a common redshift of 0.123.} 
\end{figure*}

Since the gravitational redshift of a ring of matter around the SMBH will vary inversely with radius, the line width constraint corresponds to a ring thickness of 1\%, or 0.27 R$_{g}$ and 6$\times$ 10$^{12}$ cm
for \pg. With a column density of N$_{H}$ $\sim$ 8$\times$ 10$^{19}$ cm$^{-2}$ from photionization modelling of RGS data, we find a mean density in the absorbing ring of
n$_{H}$ $\sim$ 1.3$\times$10$^{7}$ cm$^{-3}$.

For the pn data, covering a higher spectral energy range than the RGS, and with a higher mean redshift of 0.147, we again find no intrinsic line broadening, but the depth constraint is much weaker due to the
limited spectral resolution of the pn detector. However, the lower section of figure 2 does rule out a continuous inflow, while allowing thicker (and more massive rings) at both redshifts.  

In summary, it appears that while `narrow' spectral lines in both RGS and pn spectra rule out both redshift components being part of a continuous inflow, the data do allow strong gravity as the origin of the
persistent low-redshift absorption in \pg, with physically separate rings of matter around the black hole. 
As noted in the Introduction, such transient accreting rings could be a consequence of the random way in which AGN accrete, with inter-cloud shocks removing much of the incident angular momentum and allowing
matter to fall to a much smaller radius from where it may subsequently be accreted.

The launch of a powerful new wind during the final hours of the 2014 \xmm\ observation (Pounds and Nayakshin, 2024) offers strong support for that outcome, since the observed launch velocity
of 0.27c corresponds to the escape velocity at 25 R$_{g}$ from the SMBH, close to that of the ring of matter of redshift 0.123 identified here by its gravitational redshift.

\section{Conclusion} Underlying absorption is detected at redshifts of 0.123 and 0.147 in both RGS and pn spectra throughout the 7-orbit 2014 observation of \pg. A relativistic Doppler interpretation of
the lower redshift would indicate a slow-moving inflow in the outer regions of the sphere of influence of the SMBH, where a decadel accretion timescale should leave that flow still visible. However, the absence
of substantial line broadening appears to rule out a continuous flow.

In contrast, the gravitational redshift alternative provides a much better fit to the 0.147 redshift, while the launch of a powerful UFO near the end of the 2014 \xmm\ campaign, seems to have signalling
the accretion of most or all of the lower redshift ring (Pounds and Nayakshin 2024). 

The absence of any new observations of \pg\ over the past decade leaves the fate of the more massive ring at a redshift of 0.147 unknown, and archival data on UFO velocities suggest they are very often unique
and relatively short-lived events. With \xmm\ still operational, a new observation of \pg\ should be of great interest, in extending the study of a highly productive 'active' Seyfert galaxy and - strong \xmm\ legacy target. 

\section{Data Availability}
The data underlying this article are available in the XMM archive at http://nxsa.esac.esa.int/nxsa-web.

\section{Acknowledgements }
\xmm\ is a space science mission developed and operated by the European Space Agency. We acknowledge the excellent
work of ESA staff in Madrid successfully planning and conducting the \xmm\ observations, and thank James Pringle for pointing out the alternative strong gravity origin of a persistent 
low-redshift absorption.

\bsp	
\label{lastpage}

\end{document}